\documentclass[preprint,showpacs,preprintnumbers,amsmath,amssymb]{revtex4}
\usepackage[dvips]{graphicx}
\usepackage{epsfig}
\usepackage{dcolumn}
\usepackage{bm}

\begin{document}

\title{Disordered Correlated Kondo-lattice model}

\author{V.~Bryksa and W.~Nolting}

\affiliation{%
Institut f\"ur Physik, Humboldt-Universit\"at zu Berlin,
Newtonstra{\ss}e 15, D-12489 Berlin, Germany
}%


\begin{abstract}
We propose a self-consistent approximate solution of the
disordered Kondo-lattice model (KLM) to get the interconnected
electronic and magnetic properties of 'local-moment' systems
like diluted ferromagnetic semiconductors. Aiming at
$(A_{1-x}M_x)$ compounds, where magnetic (M) and non-magnetic (A)
atoms distributed randomly over a crystal lattice, we present
a theory which treats the subsystems of itinerant charge carriers
and localized magnetic moments in a homologous manner. The
coupling between the localized moments due to the itinerant
electrons (holes) is treated by a modified RKKY-theory which
maps the KLM onto an effective Heisenberg model. The exchange
integrals turn out to be functionals of the electronic selfenergy
guaranteeing selfconsistency of our theory. The disordered
electronic and magnetic moment systems are both treated by CPA-type
methods.

We discuss in detail the dependencies of the key-terms such as the
long range and oscillating effectice exchange integrals, 'the
local-moment' magnetization, the electron spin polarization, the
Curie temperature as well as the electronic and magnonic
quasiparticle densities of states on the concentration $x$ of
magnetic ions, the carrier concentration $n$, the exchange
coupling $J$, and the temperature. The shape and the effective
range of the exchange integrals turn out to be strongly
$x$-dependent. The disorder causes anomalies in the spin spectrum
especially in the low-dilution regime, which are not observed  in
the mean field approximation.
\end{abstract}

\maketitle

\section{Introduction}
There are many analytical models such as the Hubbard and the
Anderson models which are very useful for the description of real
correlated electron systems. The Kondo-lattice Model (KLM) is yet
another one. The KLM describes an interplay of itinerant
electrons in a partially filled energy band with magnetic
moments localized at certain lattice sites~\cite{Nolting0,Edw,Furukawa}. The characteristic model
properties result from an interband exchange interaction between
two well-defined subsystems: itinerant electrons and localized
spins. This model has a long history, particularly concerning 
with itinerant electron magnetism~\cite{Moriya}, heavy fermion systems~\cite{Stewart}, 
perovskite manganese oxides~\cite{Furukawa}, and so on. 

Problems connected with the substitutional disorder have recently
become more and more important in different fields of material
science, e.g., the diluted magnetic semiconductors
(DMS)~\cite{MacDonald}, the transition metal
dielectrics~\cite{Gusev}, etc. Extensive theoretical work on the problem of disorder
has been done for random Heisenberg spin systems~\cite{Harris,DveyAharon,Theumann,Theumann01,TahirKheli,Gurskii,Sherrington,Sherrington01,Fibich}.

Here, we are mainly
interested on how the disorder influences the characteristic
properties of local-moment systems such as the DMS, where magnetic
($M$) and non-magnetic ($A$) atoms are distributed randomly over a
crystal lattice ($A_{1-x}M_x$) with a given concentration of
magnetic atoms $x$. In order to answer this question different
approaches were
proposed~\cite{Theumann,Harris,Jones,Kudrnovsky,Fibich,Bouzerar01}.
More or less realistic electronic structure calculations based
on density functional theory (DFT) or  numerical
methods like classical Quantum Monte Carlo were used for
standard models like the Heisenberg spin system~\cite{Hilbert}.
However, the disadvantage of the realistic DFT-calculations come
from their strong material dependence~\cite{Kudrnovsky}. Therefore,
they do not explain the basic physics of disordered local-moment
systems in a simple way. But, it is known that a disordered KLM Hamiltonian can be mapped onto
an random Heisenberg model(shown in the text), so, in this respect better results can be obtained from
its model study.

A special challenge when treating the random Kondo-lattice model
arises with the fact that both the electron and the spin subsystem
have to be considered simultaneously and on the same level. Most
of the KLM investigations  are focused on the electronic
\cite{Takahashi,Blackman01} or
magnetic~\cite{Theumann,Harris,Fibich,Bouzerar2,Hilbert,Bouzerar01}
subsystem only. A special goal of our study is the homologous
treatment of the electronic and magnetic properties of the random
KLM, which mutually effect each other and, therefore, should be
determined self-consistently.

Nevertheless, we are forced to apply different methods to study
the influence of the disorder and dilution of the magnetic moments
subsystem on the properties of these two aforementioned
subsystems. In this paper for the itinerant electron system a proper alloy
analogy with the respective coherent potential approximation
(CPA)~\cite{Elliott} is used. For the random spin system, for
which the situation is not so clear, an equivalent ansatz must be
found. As it was done successfully for the periodic
KLM~\cite{Nolting0} ($\textit{'modified' RKKY (MRKKY)}$) one can
map the KLM-interband exchange on an effective and random
Heisenberg model. The resulting effective exchange integrals
between the localized spins will be long range and complicated
functionals of the electronic self-energy. The conventional RKKY,
resulting from second order perturbation theory is
insufficient even with a phenomenological damping
factor~\cite{Kudrnovsky,Bouzerar2}. Higher order conduction
electron self-energy effects, being taken into account by
$\textit{'modified' RKKY }$ but neglected by conventional $RKKY$,
provide the self-consistency of the full KLM. They drastically
influence the magnetic properties such as the Curie temperature.

A similar strategy of treating the disordered Kondo-lattice model has recently been used in ref.~\cite{Tang}, namely
the combination of partial solutions for the electronic and magnetic part by a modified RKKY theory.
However, the authors use different methods in order to solve the partial problems. 
For the electronic part, they apply a procedure which gives correct results only for very low band occupation. For the spin part, on the other hand, the disorder has
been solved by a mean-field ansatz. In this paper we present alternative treatments of these partial problems. 
In particular, we develop a better technique to deal with the spin disorder. 
The band electrons are discussed within CPA starting from an atomic limit alloy analogy 
without any restriction to the electron band occupation.
Several anomalies in the spin spectrum are found, which are not reproducible by the method of ref.~\cite{Tang}.

The paper is organized as follows: In Sec. 2 we describe the
methods and approximations that we have chosen to study electron and spin dynamics in
the disordered  KLM. Numerical results concerning Curie temperatures,
 the magnon densities of states and the effective exchange integrals are presented in Sec. 3
for wide range of the model parameters like exchange coupling
$J$, concentration of magnetic atoms $x$, electron band occupation
$n$, and temperature $T$. Section 4 is assigned for a summary and
an outlook.

\section{Theoretical Model}
The correlated KLM-Hamiltonian can be written in second quantized form as the sum of a kinetic energy, an exchange  interaction, and a Hubbard-type Coulomb interaction

\begin{equation}\label{total}
\hat H^K=\sum_{i,j, \sigma}t_{ij}a^{+}_{i\sigma}a_{j\sigma}+\frac{U}{2}\sum_{i\sigma}n_{i\sigma}n_{i,-\sigma}
-\frac{J}{2}\sum_{i\sigma}\left\lbrace z_\sigma S^{z}_ia^+_{i\sigma}a_{i\sigma}+S^\sigma_ia^+_{i,-\sigma} a_{i\sigma}\right\rbrace,
\end{equation}
where;
\begin{equation}
z_\sigma=\delta_{\sigma\uparrow}-\delta_{\sigma\downarrow},S^\sigma_i=S^x_i+iz_\sigma S^y_i,
\end{equation}
and $a^+_{i\sigma}\left( a_{i\sigma}\right) $ is the creation (annihilation) operator for the Wannier electron with the spin $\sigma \left( \sigma=\uparrow,\downarrow\right) $ at the site $\vec R_i$, $J$ is an exchange coupling which we assume to be positive in the following theory, $t_{ij}$ is an electron hopping integral between lattice sites,  $U$ is the Coulomb interaction, and $N$ is the total number of crystal sites. The inclusion of the Hubbard-term to the standard-KLM has only the reason to push doubly occupied levels above the Fermi level. This will be further commented on in section 2.1 where we discuss the electronic subsystem. 

In order to introduce the disorder into KLM of the $A_{1-x}M_x$ type, we introduce projection operators;
\begin{equation}\label{kon1}
X^{A}_i=
\begin{cases} 
1, & \text{\textit{- if site $i$ is $A$}} \\
0, & \text{\textit{- otherwise,}}
\end{cases}
\end{equation}
\begin{equation}\label{kon2}
X^{M}_i=
\begin{cases}
1, & \text{\textit{- if site $i$ is $M$}} \\
0, & \text{\textit{- otherwise.}}
\end{cases}
\end{equation}
Using these definitions we can change the notations in the Hamiltonian (\ref{total}) for the disorder in hopping and exchange coupling terms, respectively;
\begin{equation}
\begin{array}{c}
t_{ij}\rightarrow \sum_{k,\acute{k} \in A,M}t^{k\acute{k}}X^{k}_iX^{\acute{k}}_j, \\
J\rightarrow \sum_{k \in A,M}J^kX^{k}_i,
\end{array}
\end{equation}
where hopping parameters $t^{AA}=t^{MM}=t^{AM}=t^{MA}=t=W/6$ are the same for the different kind of atoms and are equivalent to a half band width $W$ for the case of a simple cubic lattice, and exchange couplings are $J^M=J,J^A=0$ for magnetic and nonmagnetic atoms, respectively.

After configurational averaging these operators are simply expressed by the concentration $x$ of magnetic atoms $M$.
\begin{equation}
\begin{array}{l}
\left\langle X^{A}_i\right\rangle _c=1-x,\\
\left\langle X^{M}_i\right\rangle _c=x.
\end{array}
\end{equation}

In order to study conduction-electron properties, we use the configurationally averaged single-electron Green function;
\begin{equation}
G_{ij\sigma}\left( E\right) =\overline{\left\langle \left\langle a_{i\sigma}|a^+_{j\sigma}\right\rangle \right\rangle _E}.
\end{equation}
Where the symbol $\overline{(\ldots})$ denotes the configurational ensemble average.

 The relation between the band occupation $n$ and the chemical potential $\mu$
 is as follows:
\begin{equation}
n=n_\uparrow+n_\downarrow=-\frac{1}{\pi}\sum_\sigma\int^{\infty}_{-\infty}\frac{ImG_{ii,\sigma}(E) }{e^{\beta(E-\mu)}+1}dE,
\end{equation}
with $\beta=1/kT$ is the inverse temperature.

\begin{figure}
\centerline{\includegraphics[angle=0,width=0.5\textwidth]{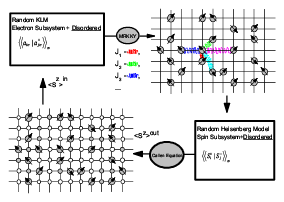}}
\bigskip
\caption{\label{fg1}General procedure of the self-consistent
mapping of the disordered Kondo-lattice model onto disordered
Heisenberg model}
\end{figure}
The main idea of this paper is to map the disordered KLM  on an effective random
isotropic spin Heisenberg Hamiltonian;
\begin{equation}
\hat H^{RH}=-\frac{1}{2}\sum_{ij}J_{ij}X^{M}_{i}X^{M}_{j}\left( S^z_iS^z_j+S^+_iS^-_j\right),
\end{equation}
where $J_{ij}$ is an effective exchange interaction between localized magnetic
moments. We assume that we can
calculate the effective exchange interaction, $J_{ij}$, using the MRKKY
method~\cite{Nolting0,Nolting1,Nolting2}
\begin{equation}\label{mrkky}
J_{\vec q} =\frac{J^2}{4\pi}\int^{\infty}_{-\infty}\frac{dE}{e^{\beta(E-\mu)}+1}\frac{1}{N}\sum_{\sigma,\vec{k}}Im\left[ G^0_{\vec k}\left( E\right) G_{\vec{k}+\vec{q},\sigma}\left( E\right) \right],
\end{equation}
which in the weak-coupling limit agrees with conventional RKKY,
where $J_{\vec q}$ is the Fourier-transform of $J_{ij}$. Then we
get the full self-consistent loop for a finite temperature
calculation (see Fig.~\ref{fg1}). In order to study  magnetic
properties of the disordered KLM we calculate the configurationally averaged magnon Green function:
\begin{equation}
D_{\vec{q}}\left(E \right)=\overline{\left\langle  \left\langle
S^+_{\vec {q}}|S^-_{-\vec{q}}\right\rangle \right\rangle_E}.
\end{equation}
Using Callen equation~\cite{Callen}, it is then possible to calculate the
magnetization $\left\langle S^z\right\rangle $:
\begin{equation}\label{Callen}
\left\langle S^z\right\rangle =\frac{\left( S-\phi\right) \left( 1+\phi\right) ^{2S+1}+\left( S+1+\phi\right) \phi^{2S+1}}{\left( 1+\phi\right) ^{2S+1}-\phi^{2S+1}},
\end{equation}
where
\begin{equation}
\phi=-\frac{1}{\pi}\frac{1}{N} \sum_{\vec{q}} \int^{\infty} _{0} \frac{ImD_{\vec{q}}\left(E\right) }{e^{\beta E}-1}dE.
\end{equation}
The procedure presented in Fig.~\ref{fg1} is very general. It is clear that we have to use different approximations in
order to solve the disordered electronic and magnetic parts. Now, we discuss these approximations in more detail.

    \subsection{Electron Subsystem: Zero-bandwidth limit of the correlated KLM}

The total Hamiltonian of the correlated KLM model is given by (\ref{total}).

The zero-bandwidth limit~\cite{Nolting2} is defined by
\begin{equation}
t_{ij} \rightarrow 0.
\end{equation}
In this approximation the excitation spectrum consists of the
following four poles~\cite{Nolting2}
\begin{equation}\label{aalen}
\begin{array}{ll} \epsilon_{1}=-\frac{1}{2}JS, & \epsilon_{2}=\frac{1}{2}J\left( S+1\right),  \\
\epsilon_{3}=U+\frac{1}{2}JS, & \epsilon_{4}=U-\frac{1}{2}J\left( S+1\right).  \end{array}
\end{equation}
It means that the single electron spectral density $A_{\sigma}\left( E\right) $ must be a four-pole function
\begin{equation}
A_{\sigma}\left( E\right)=\sum_{m=1}^4\alpha_{m\sigma}\delta\left( E-\epsilon_{m}\right).
\end{equation}
The temperature- and concentration-dependent coefficients
$\alpha_{m\sigma}$ have the physical meaning of spectral weights
for the corresponding excitation energies. The expressions for
these weight-factors are~\cite{Nolting2}:
\begin{equation}
\begin{array}{l}\label{weighte}
\alpha_{1\sigma}=\frac{1}{2S+1}\left( S+1+z_\sigma\left\langle S^z\right\rangle -\left( S+1\right) n_{-\sigma}+\Delta_{-\sigma}\right), \\
\alpha_{2\sigma}=\frac{1}{2S+1}\left( S-z_\sigma\left\langle S^z\right\rangle -Sn_{-\sigma}-\Delta_{-\sigma}\right), \\
\alpha_{3\sigma}=\frac{1}{2S+1}\left( \left( S+1\right) n_{-\sigma}+\Delta_{-\sigma}\right), \\
\alpha_{4\sigma}=\frac{1}{2S+1}\left( S n_{-\sigma}-\Delta_{-\sigma}\right),
\end{array}
\end{equation}
where $\Delta_\sigma=\left\langle
S^{\sigma}a_{-\sigma}^+a_{\sigma}\right\rangle
+z_{\sigma}\left\langle S^zn_\sigma\right\rangle $ is a mixed
spin-electron correlation function.

The term $ \Delta_\sigma$ can be expressed by the single-electron Green function~\cite{Nolting1}:
\begin{equation}
\Delta_\sigma=-\frac{1}{\pi J}\frac{1}{N}\sum_{\vec{k}}\int^{\infty}_{-\infty}\frac{dE}{e^{\beta\left( E-\mu\right) }+1}\left( E-t_{\vec{k}} \right) ImG_{\vec{k}\sigma}\left( E\right).
\end{equation}
A propagating $\sigma$ electron will meet at a certain lattice site $\vec{R_i}$ the atomic level $\epsilon_1$ with probability $\alpha_{1\sigma}$, the level $\epsilon_2$ with probability $\alpha_{2\sigma}$ and so on, if there is no correlation between sites.
 This leads to the four-component alloy. Eq.~(\ref{aalen}) makes clear that without the Coulomb interaction $U$ the pole $\epsilon_4$ would be the lowest energy. According to the spectral weight $\alpha_{4\sigma}$ in eq.~(\ref{weighte}), however, $\epsilon_4$ requires a double occupancy of the lattice site. $\epsilon_4$ to be the lowest energy therefore appears unphysical.On the other hand, for strong enough U it will not play any role.

It is easy to generalize this alloy analogy to a disordered KLM.
We have to take into consideration 
the non-magnetic sites $\epsilon_5=\epsilon_A$ as the fifth alloy constituent with the spectral
weight $1-x$. The excitation spectrum is then:
\begin{equation}
\begin{array}{lll}
\epsilon_m & \rightarrow &\tilde\alpha_{m\sigma}=x\alpha_{m\sigma}, \\
\epsilon_5=0 & \rightarrow & \tilde\alpha_{5\sigma}=1-x.
\end{array}
\end{equation}
This zero-bandwidth alloy analogy is the basic framework for applying CPA to get the electronic selfenergy.
\begin{equation}\label{atom}
\Sigma_\sigma(E)=\sum_{m=1}^{5}\tilde\alpha_{m\sigma}\frac{\epsilon_m}{1-(\epsilon_m-\Sigma_\sigma(E))G_0(E-\Sigma_\sigma(E))},
\end{equation}
where 
\begin{equation}
G_0(E)=\frac{1}{N}\sum_{\vec{k}}\frac{1}{E-t_{\vec{k}}}.
\end{equation}
\begin{figure}
\centerline{\includegraphics[angle=0,width=0.5\textwidth]{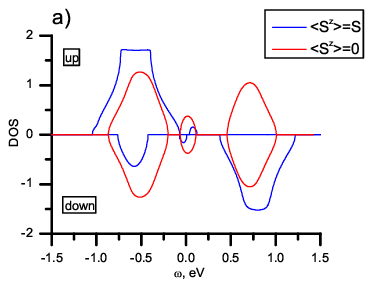}\includegraphics[angle=0,width=0.5\textwidth]{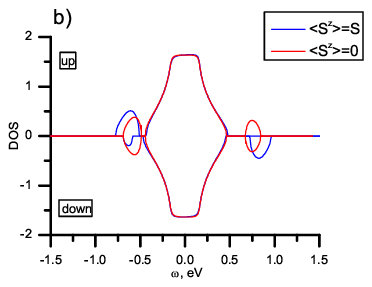}}
\bigskip
\caption{\label{fg2} Density of electron states for the disordered
KLM at a)$x=0.95$;b)$x=0.1, U \rightarrow
\infty,W=0.5eV, n=0.06, S=5/2, J=0.4eV$ for ferromagnetic $\left\langle
S^z\right\rangle =S$ and paramagnetic $\left\langle
S^z\right\rangle =0$ cases}
\end{figure}
Therewith we can write the electron Green function:

\begin{equation}
G_{\vec{k},\sigma}(E)=\frac{1}{E-\Sigma_{\sigma}(E)-t_{\vec{k}}},
\end{equation}

where $t_{\vec{k}}$ is the Fourier transform of the hopping integral $t_{ij}$. 
The quasiparticle density of states (DOS) is derived essentially from imaginary part of the Green function:
\begin{equation}
\rho_\sigma(E)=-\frac{1}{\pi N}\sum_{\vec{k}}\textit{Im} G_{\vec{k},\sigma}(E).
\end{equation}

Fig.~$\ref{fg2}$ represents the quasiparticle density of states for
 large and small concentrations x of M atoms in
full saturation $\left\langle S^z\right\rangle =S$ and
paramagnetic $\left\langle S^z\right\rangle =0$ limits, respectively. In the
case of strong Coulomb interaction($U\rightarrow \infty$), the DOS
consists in general of three subbands. For $\left\langle  S^z\right\rangle =S(T=0)$ the spin up electron density is absent for energies around
$J(S+1)/2$, while the spin down density is finite there.

In the limit $x=0$, our model reduces to the well-known Hubbard model. It has been shown~\cite{Vollhardt} that in infinite lattice dimension the
CPA is an exact treatment of the alloy problem. But, the question, however, is: what is the optimal alloy analogy for the Hubbard model?
The conventional atomic-limit starting point, which we have applied, may be questionable for the pure  Hubbard model. It is known (Ref.~\cite{Schneider}) that this alloy analogy does not allow for describing band-ferromagnetism. But, since we are interested only in the ferromagnetism due to the interband exchange $J$, we believe that the alloy analogy (Eq.~14-18) is appropriate for the correlated Kondo-lattice model although the Hubbard limit can be improved, for e.g. by the modified alloy analogy~\cite{Potthoff}.

    \subsection{Spin Subsystem}
Let us now consider a structurally disordered system of $N_M=xN$ spins which can be described by the isotropic Heisenberg Hamiltonian
\begin{equation}
\hat H=-\sum_{ij}J_{ij}X^M_iX^M_j\left( S^z_iS^z_j+S^+_iS^-_j\right).
\end{equation}

We use two-time temperature Green functions for
the investigation of spin excitations. The Green function within
Tyablikov approximation satisfies the equation of motion
\begin{equation}\label{mag1}
\begin{array}{l}
E\left\langle \left\langle S_l^{+}|S_k^{-}\right\rangle \right\rangle _E=2\delta_{lk}X^M_l\left\langle S^z_l\right\rangle+ \\
2\sum_{j\neq l}J_{lj}\left\langle \left\langle X^M_j\left\langle S^z_j\right\rangle S_l^+ -\left\langle S^z_l\right\rangle X^M_lS_j^+|S_k^-\right\rangle \right\rangle _E.
\end{array}
\end{equation}

After Fourier transformation to wave vectors, the above equation of motion can be written as:
\begin{equation}
\begin{array}{l}
\left( E-x\left\langle S^z\right\rangle E_0(\vec{q})\right) \left\langle \left\langle S_{\vec{q}}^+ |S_{\vec{q'}}^-\right\rangle \right\rangle _E=  \\
2x\left\langle S^z\right\rangle\delta(\vec{q}+\vec{q'})+2\left\langle S^z\right\rangle\frac{1}{\sqrt{N}}\Delta X_{\vec{q}+\vec{q'}}+\\
2\left\langle S^z\right\rangle\sum_{\vec{k}}\left( J(\vec{q}-\vec{k})-J(\vec{k})\right) \frac{1}{\sqrt{N}}\Delta X_{\vec{q}+\vec{k}} \left\langle \left\langle S_{\vec{k}}^+ |S_{\vec{q'}}^-\right\rangle \right\rangle_E,
\end{array}
\end{equation}
where we have used
\begin{equation}\label{deff1}
\begin{array}{c}
\Delta X_{\vec{q}}=X_{\vec{q}}-x\sqrt{N}\delta({\vec{q}}), \\
X_{\vec{q}}=\frac{1}{\sqrt{N}}\sum_{i=1}^{N}X^M_ie^{-i\vec{q}\vec{R_i}} \\
E_0(\vec{q})=2x \left\langle S^z\right\rangle (J(0)-J( \vec{q} ))
\end{array}
\end{equation}
 The approximation $\left\langle S^z\right\rangle=\overline{\left\langle S^z_l\right\rangle}$ is very often applied to disordered systems and means the neglect of structural
 fluctuations of spin positions.

Averaging over all possible realizations of atomic configurations, the equation for averaged Green function can be written in the following form:
\begin{equation}
\begin{array}{l}
\left( E-E_0(\vec{q})\right) \overline{\left\langle \left\langle S_{\vec{q}}^+ |S_{\vec{q'}}^-\right\rangle \right\rangle} _E=2x\left\langle S^z\right\rangle\delta(\vec{q}+\vec{q'})+\\
2\left\langle S^z\right\rangle\sum_{\vec{k}}\left( J(\vec{q}-\vec{k})-J(\vec{k})\right) \frac{1}{\sqrt{N}}\overline{\Delta X_{\vec{q}+\vec{k}} \left\langle \left\langle S_{\vec{k}}^+ |S_{\vec{q'}}^-\right\rangle \right\rangle}_E.
\end{array}
\end{equation}

The equation contains a higher-order averaged Green function $\overline{\Delta XG}$. One can write the equation of motion for this function, multiplying it by $\Delta X$ and performing configurational average which will include terms like, $\overline{\Delta X\Delta X G}$. In order to solve these equations, the following decoupling of configurational averages is used;

\begin{equation}\label{indep}
\overline{\Delta X_{\vec{q}-\vec{k}} \Delta X_{\vec{k}-\vec{k'}}\left\langle \left\langle S_{\vec{k'}}^+ |S_{\vec{q'}}^-\right\rangle \right\rangle}\approx\overline{\Delta X_{\vec{q}-\vec{k}} \Delta X_{\vec{k}-\vec{k'}}} \overline{\left\langle \left\langle S_{\vec{k'}}^+ |S_{\vec{q'}}^-\right\rangle \right\rangle},
\end{equation}

where
\begin{equation}
\overline{\Delta X_{\vec{q}-\vec{k}} \Delta X_{\vec{k}-\vec{k'}}}=\delta(\vec{q}-\vec{k'})\overline{\Delta X_{\vec{q}-\vec{k}} \Delta X_{\vec{k}-\vec{q}}}.
\end{equation}
The equation exploits translation symmetry.

Thus, for the averaged Green function, we eventually get;
\begin{equation}\label{gmag}
\begin{array}{l}
\overline{\left\langle \left\langle S_{\vec{q}}^+ |S_{\vec{q'}}^-\right\rangle \right\rangle_E}=\delta(\vec{q}+\vec{q'})\frac{2x\left\langle S^z\right\rangle +P(\vec{q};E)}{E-E_0(\vec{q})-\Sigma(\vec{q};E)}, \\
P(\vec{q};E)=\frac{1}{N}\sum_{\vec{k}}\frac{J(\vec{q}-\vec{k})-J(\vec{k})}{E-E_0(\vec{k})}S(\vec{q}-\vec{k}), \\
\Sigma(\vec{q};E)=\frac{1}{N}\sum_{\vec{k}}\frac{(J(\vec{q}-\vec{k})-J(\vec{q}))(J(\vec{q}-\vec{k})-J(\vec{k}))}{E-E_0(\vec{k})}S(\vec{q}-\vec{k}),
\end{array}
\end{equation}
where we have introduced  the
structure factor for the random distribution of magnetic atoms
$S(\vec{q})=[\left\langle S^z\right\rangle]^2\overline{\Delta X_{\vec{q}} \Delta
X_{-\vec{q}}}$.

\subsubsection{Virtual Crystal Approach}
This is the simplest approximation for the magnon Green function.
If we neglect all scattering processes $(P=0,\Sigma=0)$ in
Eq.~(\ref{gmag}) we obtain the following expression for the magnon
Green function:
\begin{equation}\label{vca}
\overline{\left\langle \left\langle S_{\vec{q}}^+ |S_{\vec{q'}}^-\right\rangle \right\rangle_E}=\delta(\vec{q}+\vec{q'})\frac{2x\left\langle S^z\right\rangle}{E-E_0(\vec{q})}.
\end{equation}
\subsubsection{Low Quadratic Approximation}
The next simple approach for the structure factor $S(\vec{k})$, while including some scattering processes $(P\neq 0,\Sigma\neq 0)$, is obtained
using a cumulant method~\cite{EdwardsCC,Matsubara}.
Using the definition~(\ref{deff1}) of $X_{\vec{k}}$ variables
\begin{equation}
\overline{\Delta X_{\vec{k}} \Delta X_{-\vec{k}}}=\frac{1}{N} \sum_{i,j}e^{\vec{k}(\vec{R_i}-\vec{R_j})}(\overline{X^M_i X^M_j}-x^2),
\end{equation}
and the cumulant formula for finding the average of the product of $X^M_i$ variables~\cite{EdwardsCC,Matsubara}.
\begin{equation}
\overline{X^M_i X^M_j}=\delta_{ij}P_2(x)+[P_1(x)]^2.
\end{equation}
where $P_2(x)=x(1-x),P_1(x)=x$ are second and first cumulants, respectively.
Finally, using the following approximation for the structure factor $S(\vec{k})$,
\begin{equation}\label{str1}
S(\vec{k})\rightarrow S^{eff}=[\left\langle S^z\right\rangle]^2(1-x)x,
\end{equation}
we get the expression for the magnon Green function in the Low
Quadratic Approximation(LQA) [Ref.~\cite{Fibich}].
\begin{equation}
\begin{array}{l}\label{lqa}
\overline{\left\langle \left\langle S_{\vec{q}}^+ |S_{\vec{q'}}^-\right\rangle \right\rangle_E}=\delta(\vec{q}+\vec{q'})\frac{2x\left\langle S^z\right\rangle+S^{eff}\frac{1}{N}\sum_{\vec{k}}P(\vec{k};E)}{E-E_0(\vec{q})-S^{eff}\frac{1}{N}\sum_{\vec{k}}Q(\vec{k};E)}, \\
P(\vec{k};E)=\frac{J(\vec{k}-\vec{q})-J(\vec{k})}{E-E_0(\vec{k})} \\
Q(\vec{k};E)=\frac{(J(\vec{k}-\vec{q})-J(\vec{k}))(J(\vec{k}-\vec{q})-J(\vec{q}))}{E-E_0(\vec{k})}.
\end{array}
\end{equation}

\section{Results}\label{Results}

Let us discuss some of the results obtained for finite temperature using the
self-consistent calculation according to the procedure as sketched in Fig.~\ref{fg1}.

\begin{figure}
\centerline{\includegraphics[angle=0,width=0.55\textwidth,height=16cm]{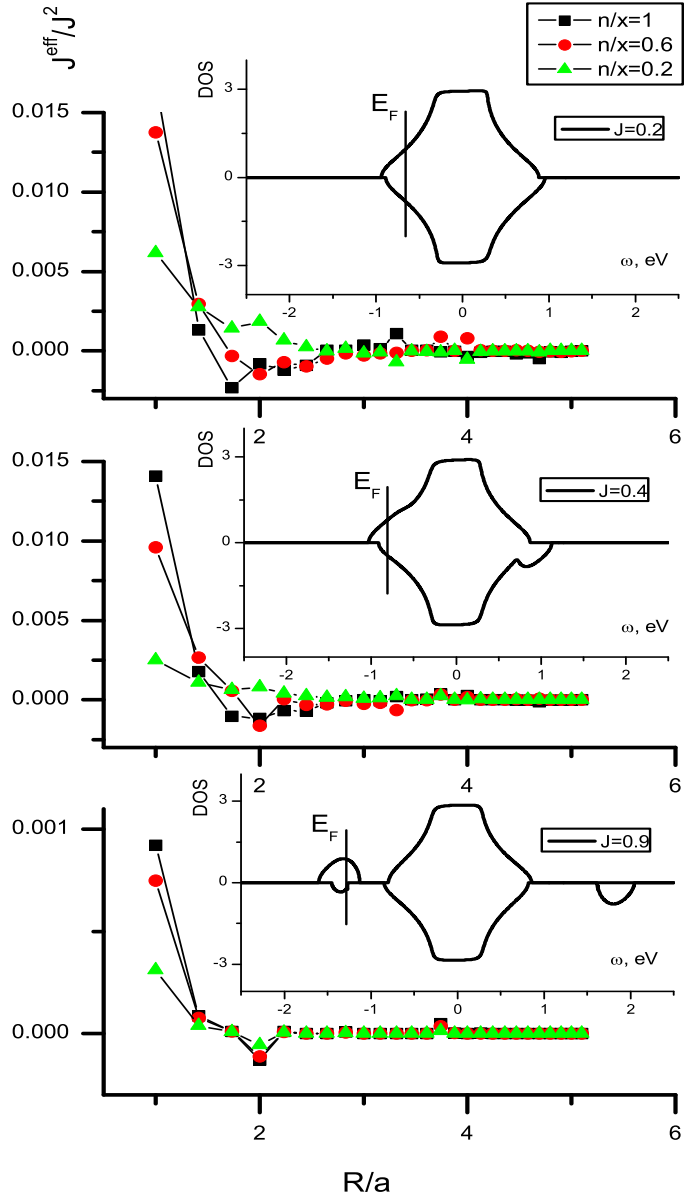}}
\bigskip
\caption{\label{fg6}Dependence of the MRKKY interaction
(Eq.~$\ref{mrkky}$) on the distance $R$ between magnetic atoms in
a simple cubic lattice ($R=1$ is a first shell(nearest-neighbors
atoms),$R=\sqrt{2}$ is a second shell(next nearest-neighbors atoms)
and so on) at fixed parameters $U \rightarrow \infty,W=0.9eV, S=5/2, x=0.1, T=0,
\left\langle S^z \right\rangle =S$ of the model for three different
coupling $J=0.2eV, J=0.4eV, J=0.9eV$ and three different values of a small
electron concentration $n/x=1,n/x=0.6, n/x=0.2$. In the inset is the electron DOS for the same
parameters. The Fermi edge is indicated for the band occupation
$n/x=0.6$.}
\end{figure}

Fig.~\ref{fg6} represents the distance-dependence of the effective (MRKKY) exchange integrals
for different values of the interband exchange parameter $J$ of the Kondo-lattice model.
For the concentrated
case ($x=1$) and in the low coupling regime $J\rightarrow 0$ the MRKKY
interaction agrees with the conventional RKKY~\cite{Nolting0,Nolting3}.
 In this situation the effective exchange constant has a
long-range character with strong oscillations in the real space.
However, with increasing $J$ the MRKKY interaction looses the long
range character and transforms into a fairly short-range
interaction (like double exchange), where only the first few effective
exchange parameters~\cite{Nolting3} turn out to be important. We
realize the same behavior for diluted systems $x < 1$. For
better comparison with the conventional RKKY, Fig.~\ref{fg6} shows only
the oscillation part of the MRKKY interaction $J^{eff}/J^2$.
The conventional RKKY interaction is a continuous function of
distance $R$, while the MRKKY approach can provide only  discrete
results. The magnetic neighbors of a given magnetic ion are
considered as ordered in $\textit{'shells'}$. The larger the shell
number L, the larger is the distance from the given magnetic ion. The
L-th shell is built up by the L-th nearest neighbors. Each
point in Fig.~\ref{fg6} corresponds to a certain shell
demonstrating the distance-dependence of the effective exchange
parameter $J^{eff}_{ij}$. Generally, there are two contributions
to the effective exchange interaction, namely a Kondo-scattering
of electrons $J$ by magnetic impurities and in addition the
scattering of the electrons by the random distribution of magnetic
atoms with the concentration $x$. It is not possible to separate
these two contributions since both of them play an equvally important role.
\begin{figure}
\centerline{\includegraphics[angle=0,width=0.9\textwidth,height=7cm]{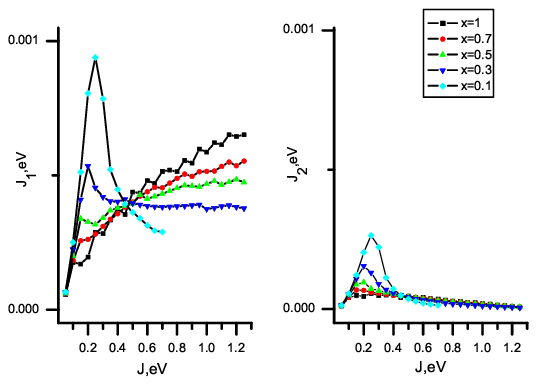}}
\bigskip
\caption{\label{fg7}Dependence of the nearest-neighbor effective exchange integral $J_1$ and the
next-nearest-neighboring one $J_2$ on the
exchange coupling $J$ at $U \rightarrow
\infty,W=0.5eV,n=0.06,S=5/2,T=0,$ for different values of
concentration $x$. }
\end{figure}
\begin{figure}[!h]
\centerline{\includegraphics[angle=0,width=0.9\textwidth,height=7cm]{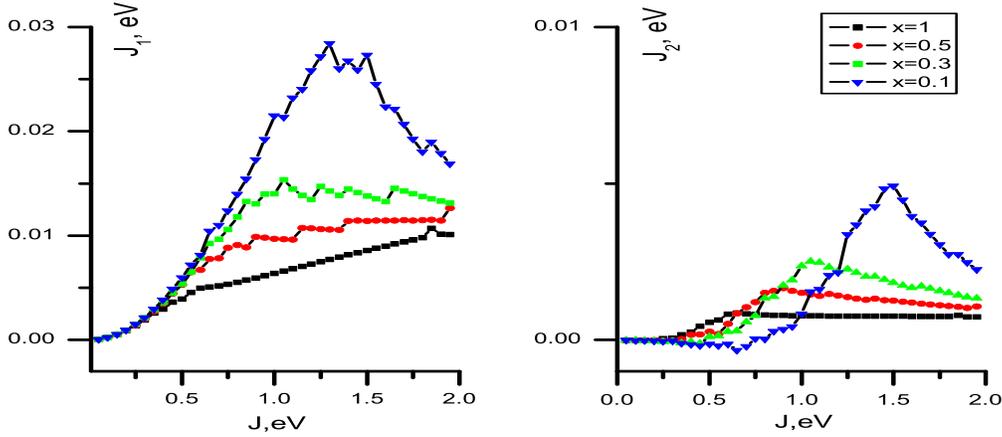}}
\bigskip
\caption{\label{fg8}The same as in Fig.~\ref{fg7}, but for $S=1/2$.}
\end{figure}

Figs.~\ref{fg7} and \ref{fg8} represent the dependencies of the
nearest-neighbor and the next-nearest-neighbor effective exchange
interaction ($J_1,J_2$) on the interband exchange coupling $J$,
and for different concentrations $x$ and for two different
spin values $S=5/2$ and $S=1/2$. In the strong coupling regime ($J>0.8 $) of
the concentrated system ($x=1$) only the short-range interaction
($J_1$) is important, the other interactions are small
($J_2,J_3,...$)~\cite{Nolting3}. In the diluted and strong coupling case, however, the
next-neighbor interaction $J_1$ becomes smaller
(Fig.~\ref{fg7}) than for $x=1$. In the low coupling regime ($J<0.1$) the results of the
modified RKKY coincide with those of the conventional RKKY
($J_1\sim J^2$) (Fig.~\ref{fg7},\ref{fg8}) which is nearly
$x$-independent. The situation is much more complicated and
non-monotonic in the intermediate coupling regime. Dilution may
even lead to an enhanced value for the nearest-neighbor localized
moment interaction $J_1$ compared with that for $x=1$
(Fig.~\ref{fg7},\ref{fg8}).
The self-consistent calculation of the effective exchange parameters
$J_1,J_2,J_3,..$ yields only a slight temperature-dependence of order $0.01$ meV or even less.
\begin{figure}
\centerline{\includegraphics[angle=0,width=0.5\textwidth]{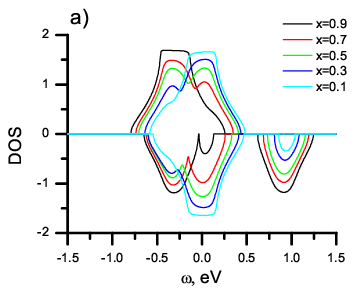}\includegraphics[angle=0,width=0.5\textwidth]{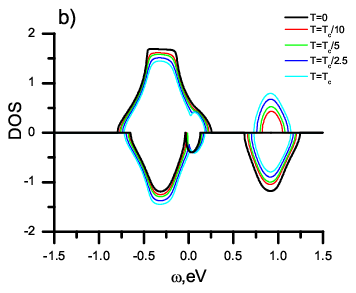}}
\bigskip
\caption{\label{fg9}Electron DOS calculated
with $U \rightarrow
\infty,W=0.5eV, J=1.2eV, S=1/2, n=0.6$ a) $T=0$ for different values of the
concentration $x$; b) $x=0.9$ for different values of the
temperature $T$.}
\end{figure}

The electronic quasiparticle-DOS for different concentrations $x$
is shown in Fig.~\ref{fg9}. There are two parts with different
physical meanings. One part consists of correlated bands centered
at $-JS/2$ and $J(S+1)/2$ due to the exchange interaction with magnetic atoms. The spectral
weights of the two subbands are strongly temperature-dependent(see Fig.~\ref{fg9}(b)). In
the case of $T=0$ the spectral weight of the upper $\uparrow$
subband disappears, but however, is present in the $\downarrow$
spectrum. The second part around $\epsilon_A=0$ represents a
non-correlated band being connected to the non-magnetic atoms (Fig.~\ref{fg2},\ref{fg9}).
This part of the spectrum is practically temperature independent.
\begin{figure}
\centerline{\includegraphics[angle=0,width=0.5\textwidth]{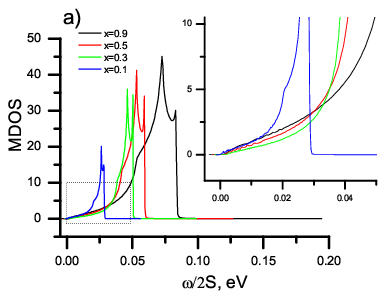}\includegraphics[angle=0,width=0.5\textwidth]{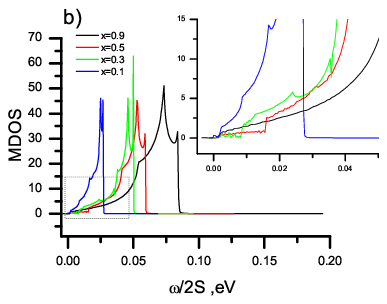}}
\bigskip
\caption{\label{fg10}Magnon DOS calculated self-consistently by use
of the VCA~(a) and the Low Quadratic Approximation~(b) with the same parameters as those in
Fig.~\ref{fg9} for different values of the concentration $x$.}
\end{figure}

In Fig.~\ref{fg10} results for the magnon density of states
are shown.
Fig.~\ref{fg10}(a) shows the magnon DOS in the VCA approximation
(\ref{vca}) for the saturated ferromagnetic ground state. We
see a typical consequence of dilution: a shift to the low energy
side with decreasing concentration $x$ of magnetic atoms. The
same is valid for the magnon DOS in the low quadratic
approximation (Eq.~\ref{lqa}) but there are some marked differences
(Fig.~\ref{fg10}(b)). Usually, in the case of weak magnon interaction (Fig.~\ref{fg10}(a)), the magnon DOS near zero energy can be expressed as
$DOS(E)=DE^2$, where $D$ is a stiffness parameter~\cite{Furukawa}. But for the strong magnon interaction, due to the disorder,
the magnon DOS near zero energy has higher contributions, like as $DOS(E)=DE^2+KE^3$, where $K$ is the disorder parameter. These contributions are
strongly modifying the magnon stiffness. For example, the huge low-energy part for low
concentrations $x$ can be observed only for strong magnon interaction (inset of Fig.~\ref{fg10}(b)). But in the VCA we have the normal $DOS(E)=DE^2$ behavior (inset of Fig.~\ref{fg10}(a)). We conclude the same as in ref.~\cite{Furukawa}, that the disorder is strongly influencing the low energy magnon DOS.
\begin{figure}
\centerline{\includegraphics[angle=0,width=0.5\textwidth]{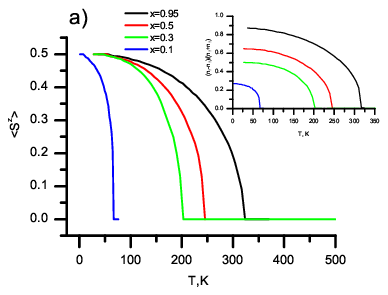}\includegraphics[angle=0,width=0.5\textwidth]{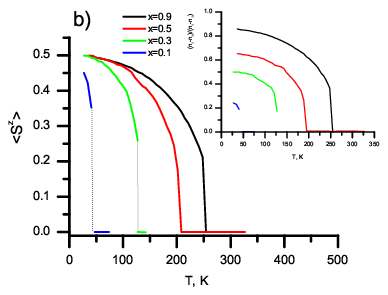}}
\bigskip
\caption{\label{fg12}Self-consistent calculation of the
dependence of the magnetization $\left\langle S^z\right\rangle $
on temperature by use of the VCA~(a) and the Low Quadratic Approximation~(b) with the same
parameters as those in Fig.~\ref{fg9} for different values of
the concentration $x$.}
\end{figure}

After the self-consistent calculation for
finite temperatures, we obtain magnetization and the resulting value of the
Curie temperature (Fig.~\ref{fg12}(a), Fig.~\ref{fg12}(b)). In
Fig.~\ref{fg12}(a), the magnetization curves are plotted
which are obtained within the VCA for the magnon subsystem
(Eq.~\ref{vca}). We see that the Curie temperature decreases with
increasing dilution $1-x$. The same holds for the electron
polarization (inset of Fig.~\ref{fg12}(a)):
$(n_\uparrow-n_\downarrow)/n$.

The magnetization curves in Fig.~\ref{fg12}(b) are determined within the low
quadratic approach (Eq.~\ref{str1},\ref{lqa}) for the
same parameters as for VCA in Fig.~\ref{fg12}(a). Since this approach
includes magnon scattering processes more realistically than the
VCA, the Curie temperature for the same concentration $x$ of
magnetic ions is lesser than that of VCA. For small $x$, the
low quadratic approximation predicts first order
ferromagnetic-paramagnetic transitions~\cite{Gusev} (dotted lines
in Fig.~\ref{fg12}(b)).

The explanation of this fact is based on predictions of the different approximations for the low energy part of the magnon
DOS (see Fig.~\ref{fg10}(a),\ref{fg10}(b)). There was no report  on first order transition in Ref.~\cite{Tang}. In our opinion this is 
related to the fact that spin  disorder was treated in mean-field like approach, which is evidently comparable to our VCA results(see Fig.~\ref{fg12}(a)), and which does not exhibit first order transition.

\section{Discussion and Summary}

There are many real materials where disorder plays an important
role for electronic as well as magnetic properties (binary
substitutional alloys, diluted magnetic semiconductors, perovskite
manganese oxides, spin glass materials, transition metal
dielectrics, etc.). Of course, these real systems are much
more complicated than  what the simple Kondo-lattice model predicts
(complicated crystal lattice, multi-band structure, hybridization
effects, spin-orbit coupling). However, we believe that the main
microscopic mechanisms are well described in terms of the
characteristic KLM features. The final goal is to make
a quantitative description of those materials combining the present
analytic model investigations with realistic $\textit{'ab
initio'}$ calculations of the band structure as it was done
previously for concentrated local-moment
systems~\cite{Hilbert,Muller}.

In this paper we discussed the influence of moment disorder on the
electron and spin excitation spectra of the random KLM. Starting
from an alloy analogy based on the exactly known zero-bandwidth
limit of the KLM we applied a CPA procedure to find out the
reaction of the electronic spectrum on the random mixture of
magnetic and nonmagnetic atoms~\cite{Nolting2}. The Hubbard term in the model Hamiltonian~(\ref{total}) helps to prevent states, which belong to double occupancies of lattice
sites, to be ground states. The analytical
expression for the electronic selfenergy has been used then to get
the effective exchange integrals of the modified RKKY theory. The
latter results from a mapping of the interband exchange onto an effective
random Heisenberg model (Fig.~\ref{fg1}) which was subsequently
treated in the spirit of the well-known Tyablikov approximation.
The disorder in the localized spin system turned out to be the most
involved part of our study (Fig.~\ref{fg1}). It was incorporated
via the equation of motion method and the technique of configurational averaging.
In order to decouple the higher-order averaged Green functions we used
the approximation of independent fluctuations~(\ref{indep}). The expression~(\ref{gmag}) for the averaged magnon Green function
is generalized by using the structure factor of disordered distribution of magnetic atoms over a crystal lattice.
Here we also used an approximation~(\ref{str1}), identical to the low quadratic approach~\cite{Fibich}.
 
There is no direct interaction between the localized moments.
Therefore, the collective order is caused by the indirect
 exchange interaction mediated by the itinerant band electrons.
Consequently, the indirect momentum coupling strongly depends on
electronic model parameters such as exchange coupling $J$ and band
occupation $n$ (see Fig.~\ref{fg6}). A further important parameter
is of course the concentration of magnetic atoms $x$. The
interactions found by modified RKKY resemble to those of the
conventional RKKY only in the low coupling limit ($J \sim 0$). In
the large coupling regime the long-range interaction
transforms into short-range interaction (see
Fig.~\ref{fg7},\ref{fg8}), where only the nearest-neighbor
interaction is important($J_1\neq 0, J_2\approx 0, J_3\approx 0, ...$). But
in the intermediate couplings we found that (
Fig.~\ref{fg7},\ref{fg8}) the effective exchange interaction
between the localized magnetic moments is strongly non-linear for
small concentration $x$ of magnetic atoms. This effect was also
considered in ref.~\cite{Blackman,DeGennes}, and also recently
discussed in ref.~\cite{Singh1,Singh2,Kudrnovsky} for diluted magnetic semiconductors.
 In ref.~\cite{Kudrnovsky,Sarma} the influence of disorder
on the RKKY interaction of two magnetic impurities was accounted for by a phenomenological
damping factor. In general, this damping factor has to be calculated
within the full Kondo-lattice model.

Another important finding is an influence of the disorder on the
magnon excitations for small concentrations of magnetic
atoms (Fig.~\ref{fg10}). We found rather different results for the
magnetic excitations in, respectively, VCA (Eq.~\ref{vca},
Fig.~\ref{fg10}(a)) and the low quadratic approach~\cite{Fibich}
(Eq.~\ref{lqa}, Fig.~\ref{fg10}(b)). It is clear that VCA yields too
simple expressions for the magnon Green's function (Eq.~\ref{lqa}
and Eq.~\ref{gmag}). Recently some treatments were proposed where the
Kondo-lattice model is considered in the mean-field (VCA)
approximation in order to explain real material
properties~\cite{Kudrnovsky,Bouzerar1,Bouzerar2} taking into
account only the dilution, while our results show that for small
concentrations the disorder effects are also very important. Such
effects play a crucial role for the understanding and
controlling key-properties such as the  Curie temperature. We
see that disorder changes the magnon-DOS  very drastically in
particular for low energy excitations~(Fig.~\ref{fg10}(b)) which are
decisive for the resulting values of the Curie
temperature~\cite{Furukawa,Gusev}.

We also note that there is another method to calculate the exchange interactions in ferromagnetic metals and alloys~\cite{Liechtenstein}.
This method~\cite{Liechtenstein,Bruno} works in the limit of infinite magnon wavelength. The MRKKY interaction does not have such a constraint. 
However, it is possible that for the disordered case the MRKKY theory (Eq.~\ref{mrkky}) has to be improved. The same leads true correct for the Callen equation (Eq.~\ref{Callen}).

Some factors have not been included in our self-consistent
calculation, for example,  the
environmental cluster effects for electron excitations. We are aware that this
effect can modify the magnetic properties. These are currently being under investigation. Furthermore, as mentioned at the end of Sect.~2.1, 
the incorporation of the Hubbard type Coulomb interaction can be improved. This will be done in a forthcoming investigation which aims at the competition between band-magnetism due to the Hubbard interaction and local-moment magnetism due to the interband exchange $J$ of the Kondo-lattice model. The atomic-limit alloy analogy is then of course not an appropriate starting point.

In this paper we have restricted  our considerations to the ferromagnetic ($J>0$) KLM, the standard ``Kondo physics'' ($J<0$)~\cite{Burdin,Mydosh} is therefore excluded from the very beginning. The main goal was to work out the interplay between disorder and magnetic stability in diluted local-moment systems. We are aware that parts of our theory have to be refined to reproduce the  standard ``Kondo Physics''($J<0$).

\section{Acknowledgments}

One of the authors (V.Bryksa) gratefully
acknowledges financial support by the Graduate College of the Deutsche Forschungsgemeinschaft: "Fundamentals and Functionality of Size and Interface Controlled Materials: Spin- and Optoelectronics". We also thank Dr. Tang for helpful discussions and  A. Sharma for critical reading of the manuscript.



\begin{thebibliography}{100}
\bibitem{Nolting0}
W. Nolting, S. Rex and Jaya S. Mathi, J. Phys. C 9 (1996)  1301
\bibitem{Edw}
D. Edwards, A. Green and K. Kubo, J. Phys. C 11 (1999) 2791
\bibitem{Furukawa}
Yukitoshi Motome, Nobuo Fururawa, Phys. Rev. B 71 (2005) 014446
\bibitem{Moriya}
Toru Moriya, Spin Fluctuations in Itinerant Electron Magnetism (Berlin: Springer 1985) p~240
\bibitem{Stewart}
G. R. Stewart, Rev. of Mod. Phys. 56 (1984) 755
\bibitem{MacDonald}
T. Sinova Jungwirth, J. Masek, J. Kucera, A. H. MacDonald, Rev. of Mod. Phys. 78 (2006) 809
\bibitem{Gusev}
A. I. Gusev, A. A. Rempel, A. J. Magerl,  Disordered and Order in Strongly Nonstoichiometric Compounds (Berlin: Springer 2001) p~607
\bibitem{Harris}
A. B. Harris, P. L. Leath, B. G. Nickel and R. J. Elliott, J. Phys. C 7 (1974) 1693
\bibitem{DveyAharon}
H. Dvey-Aharon, J. Phys. C 13 (1980) 197
\bibitem{Theumann}
A. Theumann, J. Phys. C 7 (1974) 2328
\bibitem{Theumann01}
A. Theumann, R. A. Tahir-Kheli, Phys. Rev. B 12 (1975) 1796
\bibitem{TahirKheli}
R. A. Tahir-Kheli, Phys. Rev. B 6 (1972) 2808
\bibitem{Gurskii}
Z. Gurskii, Condensed Matter Physics 3 (2000) 307
\bibitem{Sherrington}
R. Johnston and D. Sherrington, J. Phys. C 15 (1982) 3757
\bibitem{Sherrington01}
D. Sherrington, J. Phys. C 41 (1978) 1321
\bibitem{Fibich}
H. Dvey-Aharon and M. Fibich, Phys. Rev. B 18 (1978) 3491
\bibitem{Nolting1}
W. Nolting, G. Reddy, A. Ramakanth and D. Meyer, Phys. Rev. B 64 (2001) 155109
\bibitem{Nolting2}
W. Nolting and A. M. Oles, J. Phys. C 13 (1980) 823
\bibitem{Hilbert}
S. Hilbert and W. Nolting, Phys. Rev. B 70 (2004) 165203
\bibitem{Jones}
R. C. Jones, J. Phys. C 4 (1971) 2903
\bibitem{Vollhardt}
R. Vlaming and D. Vollhardt, Phys. Rev. B 45 (1992) 4637
\bibitem{Schneider}
J. Schneider, V. Drehal, phys. stat. sol. (b) 68 (1975) 207
\bibitem{Potthoff}
M. Potthoff, T. Herrmann, T. Wener and W. Nolting, phys. stat. sol. (b) 210 (1998) 199
\bibitem{Elliott}
R. J. Elliott, J. A. Krumhansl, P. L. Leath, Rev. of Mod. Phys. 46 (1974) 465
\bibitem{Kudrnovsky}
J. Kudrnovsky, I. Turek, V. Drchal, F. Maca, P. Weinberger and P. Bruno, Phys. Rev. B 69 (2004) 115208
\bibitem{Bouzerar01}
G. Bouzerar and P. Bruno, Phys. Rev. B 66 (2002) 014410
\bibitem{Blackman}
J. A. Blackman and R. J. Elliott, J. Phys. C 2 (1969) 1670
\bibitem{Blackman01}
J. A. Blackman and D. M.  Esterling, N. F. Berk, Phys. Rev. B 4 (1971) 2412
\bibitem{DeGennes}
P. G. De Gennes, Le Journal de Physique et le Radium 23 (1962) 630
\bibitem{Brout}
R. Brout, Phys. Rev. 115 (1959) 824
\bibitem{Callen}
H. B. Callen, Phys. Rev. 130 (1963) 890
\bibitem{Bouzerar1}
G. Bouzerar, J. Kudrnovsky and P. Bruno, Phys. Rev. B 73 (2006) 024411
\bibitem{Bouzerar2}
R. Bouzerar, G. Bouzerar and T. Ziman, Phys. Rev. B 73 (2006) 024411
\bibitem{Takahashi}
Masao Takahashi, Phys. Rev. B 70 (2004) 035207
\bibitem{Tang}
G. Tang, W. Nolting, Phys. Rev. B 75 (2007) 024426
\bibitem{Singh1}
Avinash Singh, Animesh Datta, K. Das Subrat and Vijay A. Singh, Phys. Rev. B 68 (2003) 235208
\bibitem{Singh2}
K. Das Subrat and Avinash Singh,  Preprint cond-mat/0506523 (2005)
\bibitem{Sarma}
D. J. Priour, S. Sarma, Phys. Rev. Lett. 97 (2006) 127201
\bibitem{Nolting3}
C. Santos and W. Nolting, Phys. Rev. B 66 (2002) 019901(E)
\bibitem{Nolting4}
W. Nolting, T. Hickel, A. Ramakanth, G. G. Reddy and M. Lipowczan, Phys. Rev. B 70 (2004) 075207
\bibitem{Muller}
W. M\"{u}ller and W. Nolting, Phys. Rev. B 66 (2002) 085205
\bibitem{EdwardsCC}
S. F. Edwards and R. C. Jones, J. Phys. C 4 (1971) 2109
\bibitem{Matsubara}
T. Matsubara and F. Yonezawa, Prog. theor. Phys. 34 (1965) 871
\bibitem{Liechtenstein}
A. I. Liechtenstein, M. I. Katsnelson, V.P. Antropov and V. A. Gubanov, J. Magn. Magn. Mater. 67 (1987) 65
\bibitem{Bruno}
M. Bruno, Phys. Rev. Lett. 90 (2003) 087205
\bibitem{Burdin}
S. Burdin and P. Fulde, Preprint cond-mat/0701598 (2007)
\bibitem{Mydosh}
J. A. Mydosh, Spin glasses: an experimental introduction (Taylor \& Francis 1993) p~256
\end{thebibliography}
\end{document}